\documentclass[aps,prb,reprint,groupedaddress]{revtex4-2}

\usepackage[colorlinks,bookmarks=false,citecolor=red,linkcolor=blue,urlcolor=blue]{hyperref}
\usepackage{amsmath}
\usepackage{mathrsfs}
\usepackage{amssymb}
\usepackage{graphicx}
\usepackage{physics}

\def\Hhat{\hat{H}}

\def\Qhat{\hat{Q}}

\def\Shat{\hat{S}}
\def\Uhat{\hat{U}}

\def\Xhat{\hat{X}}

\def\Zhat{\hat{Z}}

\def\Mbb{\mathbb{M}}

\def\rhohat{\hat{\rho}}

\def\sigmahat{\hat{\sigma}}
\def\tauhat{\hat{\tau}}

\def\zbar{\bar{z}}

\def\ketzx{\left|\{z\},\{x\}\right\rangle}
\def\brazx{\left\langle\{z\},\{x\}\right|}
\def\Ezx{E_{\{z\},\{x\}}}

\begin{document}

\title{Unfolding the Toric Code Model with Emergent Qubits}

\author{Brijesh Kumar}
\email[]{bkumar@mail.jnu.ac.in}
\affiliation{School of Physical Sciences, Jawaharlal Nehru University, New Delhi 110067}
\date{\today}

\begin{abstract}
We present the idea of emergent qubits by an exact model construction on a trestle, also generalized to arbitrary graphs.  The corresponding eigenstates are quantum paramagnetic, with free multipolar moments. We rigorously transform the toric code model on a torus, cylinder and sheet into emergent qubits, writing all the eigenstates exactly. We devise exact quantum circuits for the toric code and other eigenstates described here. The depth of the circuit for toric code eigenstates on torus grows linearly with the total number of qubits, as compared to the sublinear growth on cylinder or sheet.
\end{abstract}

\maketitle

\section{Introduction\label{sec:intro}} The toric code model presents a promising physical setting for doing fault-tolerant quantum computation~\cite{Kitaev2003,Shor1996,Girvin2023}. It is of great interest to a wide spectrum of researchers from quantum computation to condensed matter physics. It concerns interacting quantum spin-1/2's (qubits) on square lattice, made solvable by the mutually commuting four-qubit interactions, realizing topologically degenerate states. There is a keen ongoing interest in engineering the toric code and other such states experimentally~\cite{Bluvstein2022,Roushan2021,Liu2019,Song2018, Krinner2022,Zhao2022,Acharya2023}. We, in this paper, present an interesting approach to the toric code model by rigorously transforming it into independent `emergent' qubits, which leads to exact quantum circuits for producing any toric code eigenstate. 

Below, in Sec.~\ref{sec:emergentQ}, we introduce the idea of emergent qubits by an exact one-dimensional construction, analogous to the toric code,  on a trestle. A general model on arbitrary graphs is also presented. Their exact eigenstates can all be written in the matrix-product form, and are obtained by applying appropriate unitary operators on independent emergent qubits. These eigenstates are quantum paramagnetic in nature, and realize free multipolar moments. In Sec.~\ref{sec:toric-code}, we rigorously transform the toric code model into independent emergent qubits, and write all its eigenstates exactly for arbitrary interaction strengths. We do this first on a cylinder and sheet, then on torus, by constructing highly non-local unitary transformations (somewhat similar to what we did in Refs.~\cite{Kumar2009,Kumar2013}). Since these transformations reduce exactly into CNOT quantum gates, it presents us with precise quantum circuits for realizing the toric code and other eigenstates on quantum processors. The circuit for generating the toric code eigenstates on `torus' is particularly special. We describe these quantum circuits in Sec.~\ref{sec:circuits}. We conclude this paper with a summary of our findings and scope for experiments in Sec.~\ref{sec:sum}.
  
\section{Emergent Qubits on a Trestle \label{sec:emergentQ}} 
\begin{figure}[b]
   \centering
   \includegraphics[width=\columnwidth]{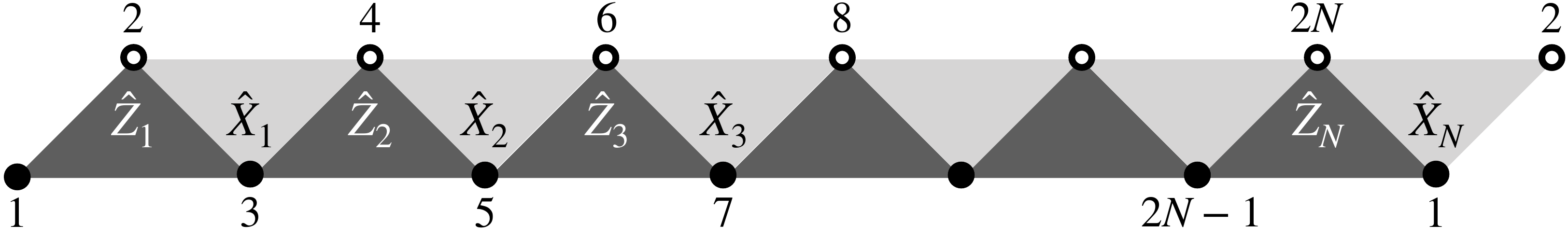} 
   \caption{Model on a closed trestle with three-qubit interactions, Eq.~\eqref{eq:model}. The dark and light gray triangles denote $\Zhat_n$ and $\Xhat_n$ operators, respectively.} 
   \label{fig:model}
\end{figure}

Consider a closed two-legged triangular strip, a trestle, of qubits interacting via the Hamiltonian given below.
\begin{equation}
\Hhat = \sum_{n=1}^N \left( I_{z,n} \Zhat_n  + I_{x,n} \Xhat_n \right)
\label{eq:model}
\end{equation}
Here, $\Zhat_n = \hat{\sigma}^z_{2n-1}\hat{\sigma}^z_{2n}\hat{\sigma}^z_{2n+1}$ and $\Xhat_n = \hat{\sigma}^x_{2n}\hat{\sigma}^x_{2n+1}\hat{\sigma}^x_{2n+2}$ are  three-qubit interactions with arbitrary strengths $I_{z,n}$ and $I_{x,n}$. See Fig.~\ref{fig:model}. The two legs of the trestle have $N$ qubits each. On a closed trestle, $\hat{\sigma}^\alpha_{2N+l}=\hat{\sigma}^\alpha_l$ for qubit labels $l=1,2,\dots,2N$ and $\alpha=z,x,y$. 
 The qubit operators $\hat{\sigma}^z_l$ and $\hat{\sigma}^x_l$ are the standard Pauli operators. The two states, $\left|\pm\right>$,  of the $l^{\rm th}$ qubit are denoted as $\left|\sigma_l\right\rangle$ for $\sigma_l=\pm1$ such that $\hat{\sigma}^z_l \left|\sigma_l\right\rangle = \sigma_l\left|\sigma_l\right\rangle$ and $\hat{\sigma}^x_l \left|\sigma_l\right\rangle = \left|\bar{\sigma}_l\right\rangle$ where $\bar{\sigma}_l=-\sigma_l$; corresponding many-qubit product states are denoted as $\left|\sigma_1,\dots,\sigma_{2N}\right\rangle = \prod^N_{n=1}\left|\sigma_{2n-1}\right\rangle\left|\sigma_{2n}\right\rangle \equiv |\{\sigma\}\rangle$. 

Note that all these $\Zhat_n$ and $\Xhat_n$ operators commute with each other, because a $Z_n$ overlaps with an $\Xhat_{n^\prime}$ by two qubits or none. Moreover, $\Zhat_n^2=\hat{1}=\Xhat_n^2$ for all $n$; here $\hat{1}$ denotes the identity operator. It implies that these operators have eigenvalues $\pm1$; let these eigenvalues be denoted as $z_n$ for $\Zhat_n$ and as $x_n$ for $\Xhat_n$. Equation~\eqref{eq:model} is a one-dimensional analog of the toric code model~\cite{Kitaev2003}. Since all these $\Zhat_n$ and $\Xhat_n$ operators commute with $\Hhat$, Eq.~\eqref{eq:model} is a system of $2N$ interacting qubits with exactly $2N$ conserved quantities. So, while its Hilbert space is $2^{2N}$ dimensional, it also has $2^{2N}$ conserved sectors given by different $z_n$'s and $x_n$'s. This ensures a one-to-one resolution of the eigenstates into conserved sectors, exhibiting perfect quantum integrability!

Let $ |\{z\},\{x\}\rangle \equiv \left|\{z_1,z_2,\dots,z_N\},\{x_1,x_2,\dots,x_N\}\right\rangle$ be an eigenstate of $\Hhat$ such that $\Zhat_n\ketzx$ $=$ $z_n\ketzx$ and $\Xhat_n\ketzx$ $=$ $x_n\ketzx$ for all $n$. Then, $\Hhat \ketzx=\Ezx\ketzx$ with eigenvalue  $\Ezx = \sum_{n=1}^N \left(I_{z,n} z_n + I_{x,n} x_n\right)$. The conserved quantum numbers $z_n$'s and $x_n$'s help us in writing the following exact expression for these eigenstates.
\begin{align}
\ketzx &=\frac{1}{2^{N/2}} \tr{\prod_{n=1}^N \Mbb^{(n)}} \label{eq:ketzx}
\end{align}
Here, $\Mbb^{(n)}$ is a matrix with elements, $\Mbb^{(n)}_{\sigma_{2n-1},\sigma_{2n+1}} = x_n^{\frac{1-\sigma_{2n+1}}{2}}  \ket{\sigma_{2n-1}}\ket{\sigma_{2n}=\sigma_{2n-1}z_n\sigma_{2n+1}}$; in matrix form, it reads as: $\Mbb^{(n)}  =    
   \begin{bmatrix} 
      |+\rangle |z_n\rangle & x_n |+\rangle |\zbar_n\rangle \\
       |-\rangle |\zbar_n\rangle & x_n |-\rangle |z_n\rangle \\
   \end{bmatrix} $.
The complete set of eigenstates given by Eq.~\eqref{eq:ketzx} presents a new many-qubit basis with $\{z\}$ and $\{x\}$ as the `emergent' qubit quantum numbers. 

We also find the unitary transformation
\begin{align}
\Uhat = \prod_{n=1}^N \left(\hat{\sigma}^z_{2n-1}\hat{\sigma}^z_{2n+1}\right)^{\Qhat^x_{2n}} = \prod_{n=1}^N \left(\hat{\sigma}^x_{2n}\hat{\sigma}^x_{2n+2}\right)^{\Qhat^z_{2n+1}} \label{eq:U}
\end{align}
that turns the interacting constituent qubits in Eq.~\eqref{eq:model} into the same number of free emergent qubits, i.e. $\Uhat^\dag \Hhat \, \Uhat$ $= \sum_{n=1}^N (I_{z,n}\hat{\sigma}^z_{2n} + I_{x,n}\hat{\sigma}^x_{2n+1})$. In Eq.~\eqref{eq:U}, $\Qhat^{\alpha}_{l}=(\hat{1}-\hat{\sigma}^\alpha_{l})/2$, and the different terms in the product mutually commute. Equation~\eqref{eq:U} connects with Eq.~\eqref{eq:ketzx} as $\ketzx = \Uhat \prod_{n=1}^N \ket{\sigma_{2n}=z_n} \frac{1}{\sqrt{2}}\sum_{\sigma_{2n+1}} x_n^{\frac{1-\sigma_{2n+1}}{2}} \ket{\sigma_{2n+1}}$. Under this $\Uhat$, the qubit operators transform as follows: 
$ \hat{\sigma}^{z(y)}_{2n} \rightarrow \hat{\sigma}^z_{2n-1}\hat{\sigma}^{z(y)}_{2n}\hat{\sigma}^z_{2n+1}$ and $ \hat{\sigma}^{x(y)}_{2n+1} \rightarrow \hat{\sigma}^x_{2n}\hat{\sigma}^{x(y)}_{2n+1}\hat{\sigma}^x_{2n+2}$, while $\hat{\sigma}^x_{2n}$ and $\hat{\sigma}^z_{2n+1}$ remain invariant.

\subsection{Properties of the emergent qubit eigenstates} 
The eigenstates, $\ketzx$, are quantum paramagnetic, as $\bra{\{z\},\{x\}} \hat{\sigma}^{\alpha}_{l} \ket{\{z\},\{x\}} = 0$ for any $l$ and $\alpha$.  But they are not spin singlets. For the total spin operator, $\Shat_\alpha =\frac{1}{2}\sum_l \sigmahat^\alpha_l$, we get $\brazx \Shat_\alpha^2 \ketzx = N/2$ for $\alpha=x,y,z$; these states have contributions from different total spins of low value ($\sim \sqrt{N}$). 
For the two-point correlations, we get $\brazx \hat{\sigma}^{\alpha_1}_{l_1} \hat{\sigma}^{\alpha_2}_{l_2} \ket{\{z\},\{x\}} = \delta_{l_1,l_2} \delta_{\alpha_1,\alpha_2}$, i.e. a complete absence of pairwise correlation between qubits! Next consider three-point correlations. Clearly, $\brazx \hat{\sigma}^z_{2n-1}\hat{\sigma}^z_{2n}\hat{\sigma}^z_{2n+1}\ketzx=z_n$ and $\brazx \hat{\sigma}^x_{2n}\hat{\sigma}^x_{2n+1}\hat{\sigma}^x_{2n+2}\ketzx = x_n$; a three-point correlation is non-zero only when the qubits form `octupolar' moments $\Zhat_n$ or $\Xhat_n$, otherwise it's zero. All sorts of correlations are zero in $\ketzx$, except those involving $\Zhat_n$'s and $\Xhat_n$'s. This is so implied by the paramagnetic form, $\Uhat^\dag \Hhat \Uhat$, of the model. These states also exhibit long-ranged correlations, first such at the six-point level: $\brazx \Zhat_{n_1} \Zhat_{n_2}\ketzx = z_{n_1}z_{n_2}$, $\brazx \Xhat_{n_1} \Xhat_{n_2}\ketzx = x_{n_1}x_{n_2}$, and $\brazx \Zhat_{n_1} \Xhat_{n_2}\ketzx = z_{n_1} x_{n_2}$ for any $n_1$ and $n_2$. More generally, $\brazx \prod_i \Zhat_{n_i} \ketzx = \prod_i z_{n_i}$ and other such forms involving $\Xhat_n$'s. But these long-ranged correlations do not represent any long-range order, because the system is paramagnetic in $\Zhat_n$'s and $\Xhat_n$'s, with no stiffness. 

The density matrix, $\hat{\rho}[\{z\},\{x\}]=\ket{\{z\},\{x\}}\bra{\{z\},\{x\}}$, of the eigenstate in Eq.~\eqref{eq:ketzx} is derived to have the following elegant form. 
\begin{equation}
\hat{\rho}[\{z\},\{x\}] =\prod_{n=1}^N\left(\frac{\hat{1}+z_n\Zhat_n}{2}\right)\left(\frac{\hat{1}+x_n\Xhat_n}{2}\right)
\end{equation} 
It is a product of the projectors for $\Zhat_n$'s and $\Xhat_n$'s with respective quantum numbers. From this, we get the reduced density matrix of a pair of qubits to be $\rhohat^{ }_{l_1,l_2} = \left(\frac{\hat{1}}{2}\right)_{l_1} \left(\frac{\hat{1}}{2}\right)_{l_2}$ for any $l_1\neq l_2$. Thus, in these eigenstates, no two qubits remain entangled upon tracing out the rest. But this tracing results in the maximally mixed state, $\frac{\hat{1}}{2}$, for each qubit~\footnote{A $D$-dimensional quantum system is said to be in a maximally mixed state if it is found in any of its $D$ basis states with equal probability $1/D$. Hence, the maximally mixed state is ${\hat{1}}/{D}$, and it has the maximum entropy $\ln{D}$.}. It means the constituent qubits in $\ketzx$ are entangled, but not pairwise. To see this collective entanglement, trace over all the even numbered qubits of the trestle. It gives the reduced density matrix, $\hat{\rho}^{ }_o= \left(\hat{1}+\prod_{n=1}^N x_n \hat{\sigma}^x_{2n-1}\right)/{2^N}$, for the remaining $N$ odd numbered qubits; likewise, the reduced density matrix of $N$ even numbered qubits is $\hat{\rho}^{ }_e= \left(\hat{1}+\prod_{n=1}^N z_n \hat{\sigma}^z_{2n}\right)/{2^N}$. Tracing out one more qubit in $\rhohat^{ }_o$ or $\rhohat^{ }_e$ gives $\left(\frac{\hat{1}}{2}\right)^{\otimes (N-1)}$, i.e. any further tracing immediately separates all the remaining qubits at once. Thus, $N$ odd-numbered qubits only collectively entangle with $N$ even-numbered qubits in these states. The entropy of this entanglement in every $\ketzx$ is $ \mathcal{S}^{ }_{o,e}=-\tr^{ }_e\{\rhohat^{ }_e \ln{\rhohat^{ }_e}\} = -\tr^{ }_o\{\rhohat^{ }_o \ln{\rhohat^{ }_o}\} =N\ln{2} - \ln{2}$. Here the entropy deficit, $-\ln{2}$, arises because while tracing over all the even (or odd) qubits, the last qubit traced (whichever that be) doesn't generate entropy (mixing); it amounts to mixing only half of the Hilbert space of all odd (or even) qubits.  

\subsection{General model on arbitrary graphs}
\begin{figure}[t]
   \centering
   \includegraphics[width=.7\columnwidth]{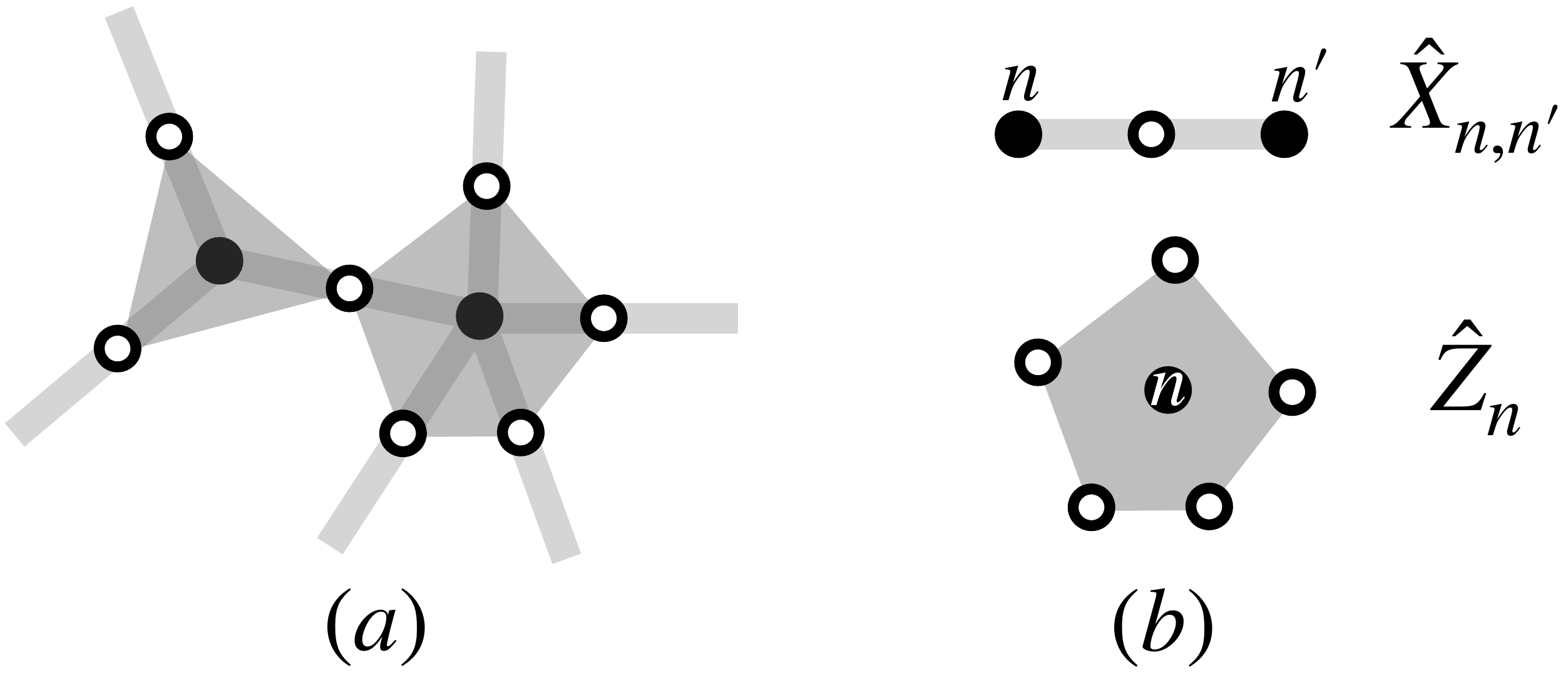} 
   \caption{$(a)$ Model with exact emergent qubit eigenstates on arbitrary graph. $(b)$ Interaction operators involving qubits sitting on sites (filled circles) and bonds (empty circles). There are no isolated sites here; every site makes a bond with at least one other site.}
   \label{fig:general-model}
\end{figure}

This construction on trestle can be easily adapted to arbitrary lattices. Consider any lattice (graph) with qubits sitting on its sites (nodes) and on the bonds connecting these sites. A bond between two nodes $n$ and $n^\prime$ is denoted as $(n,n^\prime)$; see Fig.~\ref{fig:general-model}. Define a multi-qubit interaction, $\Zhat_n = \sigmahat^z_n \prod_{n^\prime} \tauhat^z_{(n,n^\prime)}$, between the qubit at node $n$ and the qubits on all the bonds meeting at $n$; the qubit operators on the nodes and the bonds are denoted respectively as $\sigmahat^\alpha_n$ and $\tauhat^\alpha_{(n,n^\prime)}$. Also define a three-qubit interaction, $\Xhat^{ }_{(n,n^\prime)} = \sigmahat^x_n \, \tauhat^x_{(n,n^\prime)} \, \sigmahat^x_{n^\prime}$, between the qubits on a bond and the sites it connects. Now consider the Hamiltonian 
\begin{equation}
\Hhat^{ }_2 = \sum^{N_s}_{n} I_{z,n} \Zhat_n + \sum^{N_b}_{(n,n^\prime)} I_{x,(n,n^\prime)} \Xhat_{(n,n^\prime)}
\label{eq:H2}
\end{equation} 
with $N_s$ sites and $N_b$ bonds, and arbitrary interaction strengths $I_{z,n}$ and $I_{x,(n,n^\prime)}$. All these $\Zhat_n$'s and $\Xhat_{(n,n^\prime)}$'s commute with each other, and $\Zhat_n^2=\hat{1} = \Xhat_{(n,n^\prime)}^2$. Hence, $\Hhat_2$ has $N_s+N_b$ conserved $\Zhat_n$'s and $\Xhat_{(n,n^\prime)}$'s with respective quantum numbers $z_n$'s  and $x_{(n,n^\prime)}$'s of value $\pm 1$, which fully resolve the eigenstates of $\Hhat_2$. These constructions are similar to those for cluster states~\cite{Raussendorf2003,Skrovseth2009,Son2012}.

In a straightforward generalization of Eq.~\eqref{eq:ketzx}, the eigenstates of $\Hhat_2$, with eigenvalues $E_{\{z\},\{x\}} = \sum^{N_s}_n I_{z,n} z_n + \sum^{N_b}_{(n,n^\prime)} I_{x,(n,n^\prime)} x_{(n,n^\prime)}$, can be written as: 
\begin{equation}
\ketzx = \frac{1}{2^{(N_b/2)}}  \sum_{\{\sigma\}} \sum_{\{\tau\}} \mathcal{M}_{\{\sigma\},\{\tau\}} \ket{\{\sigma\},\{\tau\}}
\label{eq:ketzx-graph}
\end{equation}
where $\ket{\{\sigma\},\{\tau\}} = \prod_n^{N_s}\ket{\sigma_n}\prod_{(n,n^\prime)}^{N_b}\ket{\tau_{(n,n^\prime)}}$ are the basis states with $\sigma_n=\pm1$ and $\tau_{(n,n^\prime)}=\pm1$ as the quantum numbers of $\sigmahat^z_n$ and $\tauhat^z_{(n,n^\prime)}$ respectively, and the tensor coefficients of linear superposition are given below.  
\begin{align}
\mathcal{M}_{\{\sigma\},\{\tau\}} = \prod_{(n,n^\prime)}^{N_b} x_{(n,n^\prime)}^{\frac{1-\tau_{(n,n^\prime)}}{2}} ~ \prod_n^{N_s}\delta_{\sigma_n, z_n\prod_{n^\prime}\tau_{(n,n^\prime)}}
\label{eq:tensor}
\end{align} 
The transformation that turns $\Hhat_2$ into independent emergent qubits can be written as: 
\begin{equation}
\Uhat_2 = \prod_n^{N_s} \left[ \prod_{n^\prime} \tauhat^z_{(n,n^\prime)}\right]^{\Qhat^{x}_{n}} = \prod^{N_b}_{(n,n^\prime)}\left(\sigmahat^x_n\sigmahat^x_{n^\prime}\right)^{\Qhat^{z}_{(n,n^\prime)}}
\label{eq:U2}
\end{equation} 
where $\Qhat^{x}_{n} = \left(\hat{1}-\sigmahat^x_n\right)/2$ and $\Qhat^{z}_{(n, n^\prime)} = \left[\hat{1}-\tauhat^z_{(n,n^\prime)}\right]/2$. This generalization of Eq.~\eqref{eq:U} leads to $\Uhat_2^\dag \Hhat_2 \Uhat_2^{ } = \sum^{N_s}_n I_{z,n} \sigmahat^z_n + \sum^{N_b}_{(n,n^\prime)} I_{x,(n,n^\prime)} \tauhat^x_{(n,n^\prime)}$ and $\ketzx = \Uhat_2 \prod_n^{N_s} \prod^{N_b}_{(n,n^\prime)} \ket{\sigma_n=z_n} \frac{1}{\sqrt{2}} \sum_{\tau_{(n,n^\prime)}} x_{(n,n^\prime)}^{\frac{1-\tau_{(n,n^\prime)}}{2}} \ket{\tau_{(n,n^\prime)}}$. These states too do not exhibit dipolar (magnetic) order, but carry multi-ploar moments as the expectation values of $\Zhat_n$'s and $\Xhat_{(n,n^\prime)}$'s. Under the unitary transformation, $\Uhat_2$, we get $\sigmahat^{z(y)}\rightarrow \sigmahat^{z(y)}\prod_{n^\prime}\tau^z_{(n,n^\prime)}$ and $\tau^{x(y)}_{(n,n^\prime)}\rightarrow \sigmahat^x_n\, \tau^{x(y)}_{(n,n^\prime)}\, \sigmahat^x_{n^\prime}$, while $\sigmahat^x_n$ and $\tauhat^z_{(n,n^\prime)}$ remain invariant. The model in the presence of field, $\Hhat_2 + \sum_n h_{x,n} \sigmahat^x_n+\sum_{(n,n^\prime)} h_{z,(n,n^\prime)}\tauhat^z_{(n,n^\prime)}$, also transforms exactly into independent qubits under $\Uhat_2$; the fields $ h_{x,n}$ and $h_{z,(n,n^\prime)}$ induce magnetic moment in the eigenstates. 

\section{Unfolding the Toric Code Model\label{sec:toric-code}}  
Now we transform the toric code model into emergent qubits. The procedure in this case is more intricate but still exact. So, we first demonstrate this on a cylinder and a sheet, where it turns out to be relatively easier. Then, we do it on the torus. 

\begin{figure}[b]
   \centering
\includegraphics[width=\columnwidth]{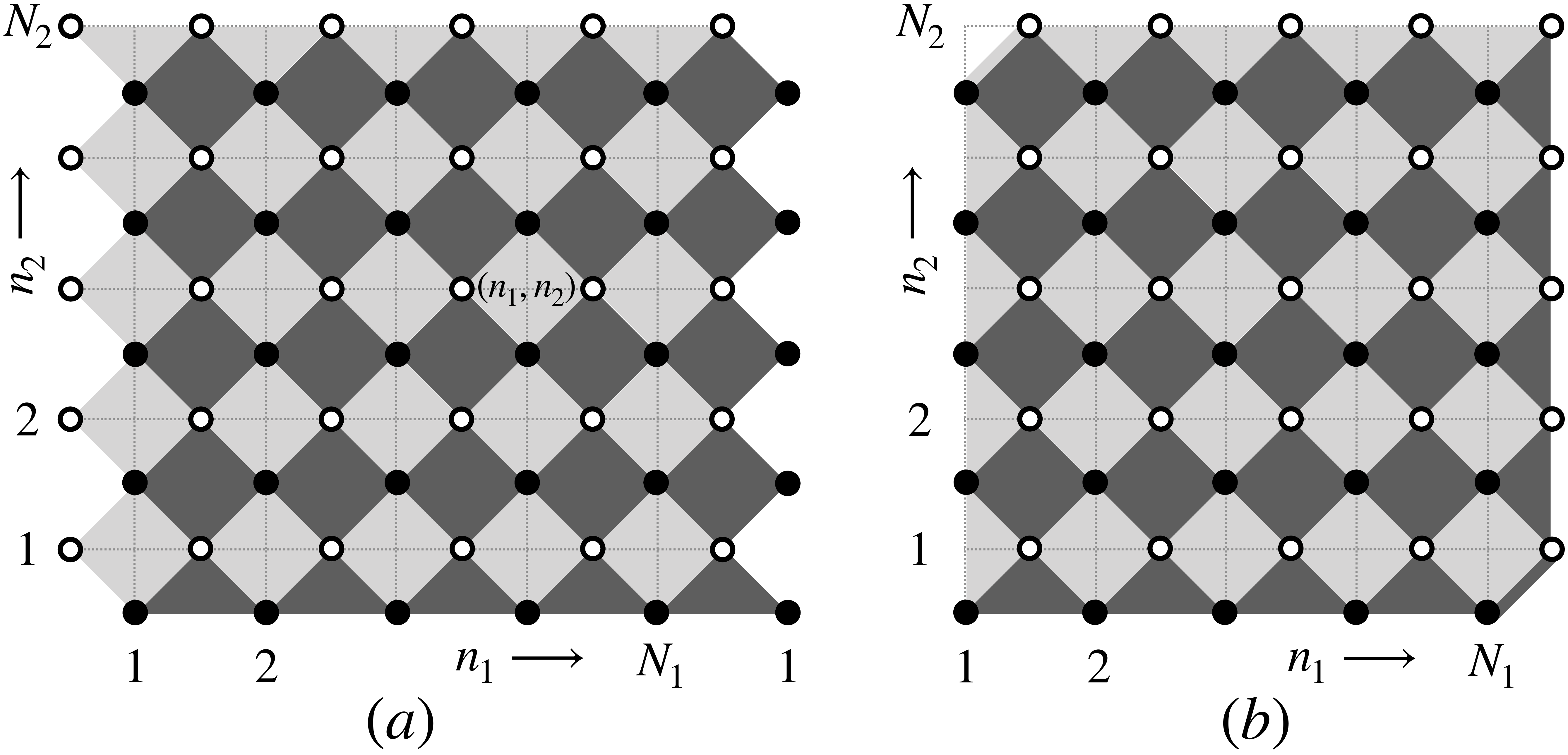}
   \caption{$(a)$ A toric code model on a cylinder; it is called $\Hhat_3$ in the text. It is periodic along $n_1$ and open along $n_2$. Dark (light) gray squares (triangles) denote $\Zhat (\Xhat) $ interactions involving four (three) qubits. $(b)$ A planar toric code, termed $\Hhat^\prime_3$. It has open boundaries along $n_1$ as well as $n_2$. Besides the four- and three-qubit interactions, it also has two two-qubit interactions of the $\Xhat$ and $\Zhat$ type at the top-left and bottom-right corners shown respectively by a light and a dark gray thick line. The total number of qubits in both cases is $2N_1N_2$.}    
   \label{fig:cylinder-sheet}
\end{figure}

\subsection{Models on cylinder and sheet}
Consider the toric code models depicted in Fig.~\ref{fig:cylinder-sheet}, obtained by cutting open the torus into a cylinder or a flat sheet, with appropriate interactions at the boundaries in addition to the standard four-qubit interactions in the bulk. On the two open ends of the cylinder, we add three-qubit interactions as in the model on trestle; see Fig.~\ref{fig:cylinder-sheet}$(a)$. On the sheet, as shown in Fig.~\ref{fig:cylinder-sheet}$(b)$, we pad the boundaries with the same three-qubit interactions, but also include two two-spin interactions at the opposite end of any one diagonal; this is an example of a planar code~\cite{Dennis2002} but with a particular boundary condition. 

Let the Hamiltonians of the models on cylinder and sheet be respectively called $\Hhat_3$ and $\Hhat^\prime_3$. They both can be written in a common form given below, but with a corresponding reading of the interaction terms at the edges. 
\begin{eqnarray}
\Hhat_{3}~({\rm or}~\Hhat^\prime_3) &=& \sum_{n_1=1}^{N_1}\sum_{n_2=1}^{N_2}\left[ I^x_{n_1,n_2} \Xhat_{n_1,n_2} + \right. \nonumber \\
&& \left. I^z_{n_1+\frac{1}{2},n_2-\frac{1}{2}} \Zhat_{n_1+\frac{1}{2},n_2-\frac{1}{2}} \right]
\label{eq:H3}
\end{eqnarray}
Here the integers $n_1$ and $n_2$ specify the sites of the square lattice formed by thin dotted lines in Fig.~\ref{fig:cylinder-sheet}. It is periodic along $n_1$ and open along $n_2$ for the cylinder in Fig.~\ref{fig:cylinder-sheet}$(a)$, and open along both $n_1$ and $n_2$ directions for the sheet in Fig.~\ref{fig:cylinder-sheet}$(b)$. The qubits in the bulk of both the models interact via the four-qubit operators, $\Xhat_{n_1,n_2} = \sigmahat^x_{n_1-\frac{1}{2},n_2} \sigmahat^x_{n_1,n_2+\frac{1}{2}} \sigmahat^x_{n_1+\frac{1}{2},n_2} \sigmahat^x_{n_1,n_2-\frac{1}{2}}$ and $\Zhat_{n_1+\frac{1}{2},n_2-\frac{1}{2}} = \sigmahat^z_{n_1,n_2-\frac{1}{2}} \sigmahat^z_{n_1+\frac{1}{2},n_2} \sigmahat^z_{n_1+1,n_2-\frac{1}{2}} \sigmahat^z_{n_1+\frac{1}{2},n_2-1} $, shown respectively by the light and dark gray squares.  

In $\Hhat_3$, the interactions at the bottom and top edges of the cylinder, denoted by dark and light gray triangles in Fig.~\ref{fig:cylinder-sheet}$(a)$, are the three-qubit interactions, $\Zhat_{n_1+\frac{1}{2},\frac{1}{2}} =  \sigmahat^z_{n_1,\frac{1}{2}} \sigmahat^z_{n_1+\frac{1}{2},1}  \sigmahat^z_{n_1+1,\frac{1}{2}} $ and $\Xhat_{n_1,N_2} = \sigmahat^x_{n_1-\frac{1}{2},N_2} \sigmahat^x_{n_1+\frac{1}{2},N_2} \sigmahat^x_{n_1,N_2-\frac{1}{2}}$ respectively. 
 
In $\Hhat^\prime_3$, the interactions on the boundary of the sheet in Fig.~\ref{fig:cylinder-sheet}$(b)$ 
read as follows: along left edge, $\Xhat_{1,n_2} =  \sigmahat^x_{1,n_2-\frac{1}{2}}\sigmahat^x_{1,n_2+\frac{1}{2}} \sigmahat^x_{\frac{3}{2},n_2}$ for $n_2=1$ to $N_2-1$; at the top-left corner, $\Xhat_{1,N_2} = \sigmahat^x_{1,N_2-\frac{1}{2}}\sigmahat^x_{\frac{3}{2},N_2}$; along the top edge, $\Xhat_{n_1,N_2}=\sigmahat^x_{n_1-\frac{1}{2},N_2}\sigmahat^x_{n_1+\frac{1}{2},N_2}\sigmahat^x_{n_1,N_2-\frac{1}{2}}$ for $n_1=2$ to $N_1$; along right edge, $\Zhat_{N_1+\frac{1}{2},n_2-\frac{1}{2}}=\sigmahat^z_{N_1+\frac{1}{2},n_2}\sigmahat^z_{N_1+\frac{1}{2},n_2-1}\sigmahat^z_{N_1,n_2-\frac{1}{2}}$ for $n_2=2$ to $N_2$; at the bottom-right corner, $\Zhat_{N_1+\frac{1}{2},\frac{1}{2}} = \sigmahat^z_{N_1+\frac{1}{2},1}\sigmahat^z_{N_1,\frac{1}{2}}$; and, along the bottom edge, $\Zhat_{n_1+\frac{1}{2},\frac{1}{2}} =  \sigmahat^z_{n_1,\frac{1}{2}}  \sigmahat^z_{n_1+1,\frac{1}{2}} \sigmahat^z_{n_1+\frac{1}{2},1} $ for $n_1=1$ to $N_1-1$. Note the two-qubit interactions, $\Xhat_{1,N_2}$ and $\Zhat_{N_1+\frac{1}{2},\frac{1}{2}}$, at the two corners in Fig.~\ref{fig:cylinder-sheet}$(b)$ denoted respectively by a light and a dark gray thick line. One could play with other choices for the boundaries, but this one is particularly simple as we will see. The interaction strengths, $I^x_{n_1,n_2}$ and $I^z_{n_1+\frac{1}{2},n_2-\frac{1}{2}}$, in these models are completely arbitrary.

We construct a unitary transformation $\Uhat_{3} = \Uhat_{3_1} \Uhat_{3_2}$ using the operators given below. It maps $\Hhat_3$ as well as $\Hhat^\prime_3$ into independent emergent qubits.
\begin{subequations}
\label{eq:U3}
\begin{eqnarray}
\Uhat_{3_1} & =& \prod_{n_1=1}^{N_1} \left[ \prod_{n_2=1}^{N_2} \left(\Xhat^{ }_{n_1,n_2} \, \sigmahat^x_{n_1,n_2-\frac{1}{2}}\right)^{\Qhat^z_{n_1,n_2-\frac{1}{2}}} \right] \label{eq:U3_1}\\
\Uhat_{3_2} &=& \prod_{n_1=1}^{N_1} \left[\prod_{n_2=N_2}^{2} \left( \sigmahat^z_{n_1+\frac{1}{2},n_2-1}\right)^{\Qhat^x_{n_1+\frac{1}{2},n_2}} \right] \label{eq:U3_2}
\end{eqnarray}
\end{subequations}
Under $\Uhat_{3_1}$, with respective $\Xhat_{n_1,n_2}$'s for $\Hhat_3$ and $\Hhat^\prime_3$, the qubits shown by filled circles in Fig.~\ref{fig:cylinder-sheet} become free, while the empty circles form independent Ising chains along $n_2$. Under $\Uhat_{3_2}$, these Ising chains also transform into independent qubits. Hence, both $\Uhat^\dag_3 \Hhat^{ }_3\Uhat^{ }_3$ and $\Uhat^\dag_3 \Hhat^{\prime}_3\Uhat^{ }_3 = \sum_{n_1=1}^{N_1}\sum_{n_2=1}^{N_2} \Big\{I^x_{n_1,n_2}  \sigmahat^x_{n_1,n_2-\frac{1}{2}} + I^z_{n_1+\frac{1}{2},n_2-\frac{1}{2}} \sigmahat^z_{n_1+\frac{1}{2},n_2}\Big\} $. Their exact eigenstates, written in the density matrix form as 
\begin{eqnarray}
\rhohat_3[\{z\},\{x\}] &=&  \prod_{n_1=1}^{N_1}\prod_{n_2=1}^{N_2} \left(\frac{\hat{1}+x_{n_1,n_2}\Xhat_{n_1,n_2}}{2}\right) \times \nonumber \\
&& \left(\frac{\hat{1}+z_{n_1+\frac{1}{2},n_2-\frac{1}{2}}\Zhat_{n_1+\frac{1}{2},n_2-\frac{1}{2}}}{2}\right), 
\label{eq:ketzx-cylinder}
\end{eqnarray} 
are given completely by the emergent qubit quantum numbers $z_{n_1+\frac{1}{2},n_2-\frac{1}{2}}=\pm1$ and $x_{n_1,n_2}=\pm1$ such that $\Xhat_{n_1,n_2}\,\rhohat_3=x_{n_1,n_2}\,\rhohat_3$ and $\Zhat_{n_1+\frac{1}{2},n_2-\frac{1}{2}}\,\rhohat_3=z_{n_1+\frac{1}{2},n_2-\frac{1}{2}}\,\rhohat_3$. The expectation values and correlations of the spin operators, except those in the form of $\Zhat$'s and $\Xhat$'s, will be zero in these states, as implied by the exact paramagnetic form of the transformed Hamiltonians. 

These models on cylinder and sheet with a particular choice of boundary interactions have uniquely resolved eigenstates because they realize the same number of emergent qubits as the constituent qubits, i.e. $2N_1N_2$. This is like the model on closed trestle, but unlike the toric code model on torus (discussed below). Other choices for the interactions on the boundary could have other consequences. For instances, an $\Hhat_3$ without the three-qubit interactions at the open ends of the cylinder will have a degeneracy of $2^{2N_1}$, or an $\Hhat^\prime_3$ without the two-qubit interactions at the two corners of the sheet will have a fourfold degeneracy, in every eigensubspace.  

\subsection{Model on torus}
\begin{figure}[t]
   \centering
   \includegraphics[width=\columnwidth]{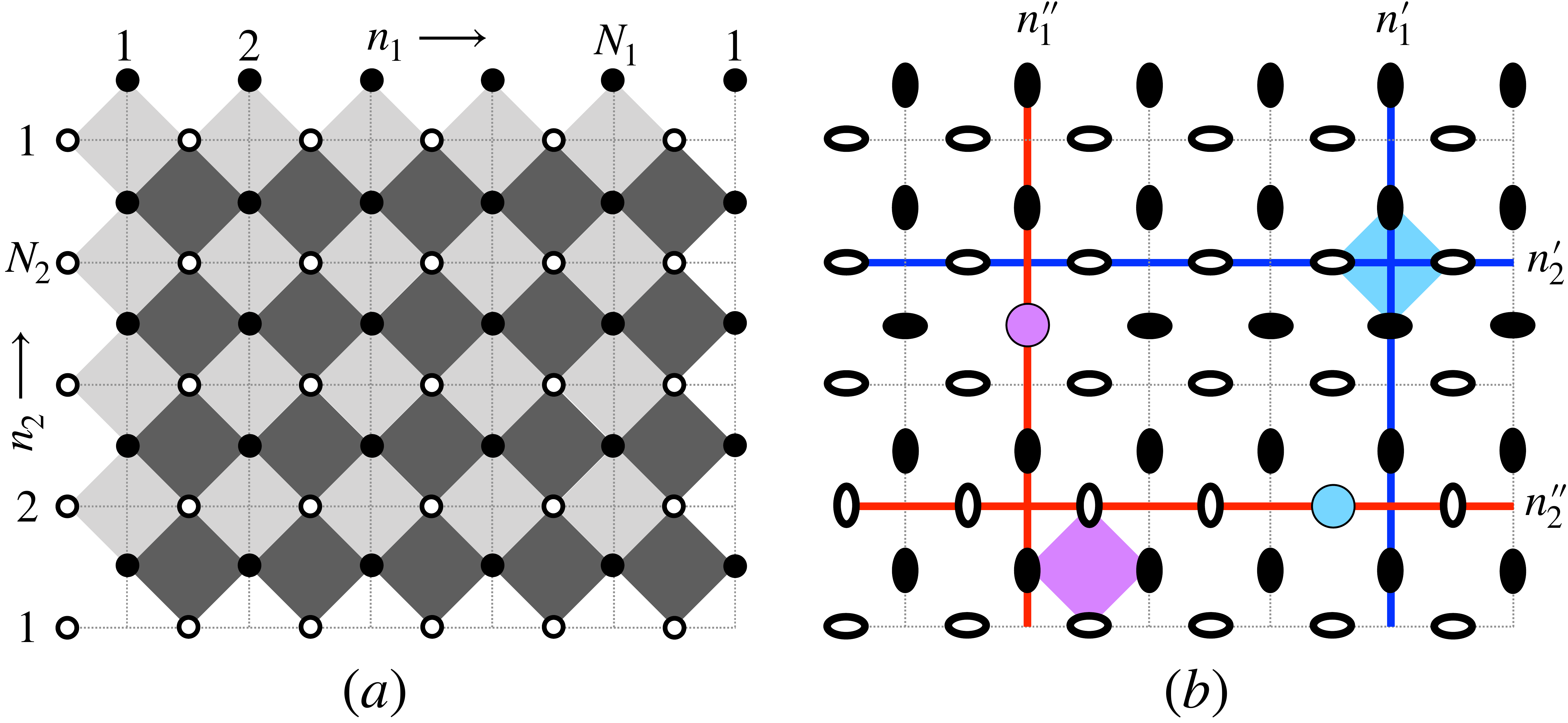}
   \caption{$(a)$ The toric code model, $\Hhat_4$. $(b)$ The transformed toric code, $\Uhat^\dag_4\Hhat_4\Uhat_4$, with $\Uhat_4$ defined in Eq.~\eqref{eq:U4}. The original interacting qubits (filled and empty circles) of $(a)$ transform into independent emergent qubits (vertical ovals denoting $\sigmahat^x$ and horizontal denoting $\sigmahat^z$) in $(b)$, and two missing qubits (colored circle on the red lines). Four (colored) reference lines at $n_{1(2)}^{\prime}$ and $n_{1(2)}^{\prime\prime}$ required for $\Uhat_4$ are arbitrary; they mark the positions of two accumulation terms (colored plaquettes) in Eq.~\eqref{eq:H4} and the two missing qubits.
   }   
   \label{fig:toric-code}
\end{figure}

Now consider the toric code model on `torus'~\cite{Kitaev2003}; see Fig.~\ref{fig:toric-code}($a$). It is periodic along $n_1$ as well as $n_2$. Let its Hamiltonian be called $\Hhat_4$, which has the same form as Eq.~\eqref{eq:H3}, but is naturally devoid of boundaries; it only has four-qubit interactions like in the bulk of $\Hhat_3$ and $\Hhat^\prime_3$. Let us devise a transformation $\Uhat_4 = \Uhat_{4_1}\Uhat_{4_2}\Uhat_{4_3}\Uhat_{4_4}$ with the following unitary operators. 
\begin{subequations}
\label{eq:U4}
\begin{eqnarray}
\Uhat_{4_1} & = & \prod_{n_1=1}^{N_1} \left[\prod_{n_2=n_2^\prime+1}^{n_2^\prime-1+N_2} \left(\Xhat^{ }_{n_1,n_2} \, \sigmahat^x_{n_1,n_2-\frac{1}{2}}\right)^{\Qhat^z_{n_1,n_2-\frac{1}{2}}} \right] \\ 
\Uhat_{4_2} & = & \prod_{n_1=1}^{N_1} \left[\prod_{n_2=n_2^{\prime\prime}-1}^{n_2^{\prime\prime}+1-N_2} \bigg(\sigmahat^z_{n_1+\frac{1}{2},n_2-1} \times \right. \nonumber \\ 
&& \left. \left[\sigmahat^z_{n_1,n_2-\frac{1}{2}}\sigmahat^z_{n_1+1,n_2-\frac{1}{2}}\right]^{\delta_{n_2,n_2^\prime}} \bigg)^{\Qhat^x_{n_1+\frac{1}{2},n_2}} \right]\\
\Uhat_{4_3} & = & \prod_{n_1=n_1^\prime+1}^{n_1^\prime-1+N_1} \left[ \sigmahat^x_{n_1+\frac{1}{2},n_2^{\prime\prime}} \prod_{n_2\neq n_2^\prime}\sigmahat^x_{n_1,n_2-\frac{1}{2}}\right]^{\Qhat^z_{n_1-\frac{1}{2},n_2^{\prime\prime}}} \\
\Uhat_{4_4} & = & \prod_{n_1=n_1^{\prime\prime}+1}^{n_1^{\prime\prime}-1+N_1} \left[\sigmahat^z_{n_1+1,n_2^\prime-\frac{1}{2}} \prod_{n_2\neq n_2^{\prime\prime}} \sigmahat^z_{n_1+\frac{1}{2},n_2}\right]^{\Qhat^x_{n_1,n_2^\prime-\frac{1}{2}}}
\end{eqnarray}
\end{subequations}
This $\Uhat_4$ is an extension of $\Uhat_3$. It is defined with reference to the two arbitrary rows (columns) labelled by integers $n_2^\prime$ and $n_2^{\prime\prime}$ ($n_1^\prime$ and $n_1^{\prime\prime}$); see Fig.~\ref{fig:toric-code}($b$).  By applying this transformation to the toric code model, we get  
\begin{align}
& \Uhat^\dag_4 \Hhat^{ }_4 \Uhat^{ }_4 = \nonumber \\
& \sum_{n_1=1}^{N_1} \left(\sum_{n_2\neq n_2^\prime} I^x_{n_1,n_2}\sigmahat^x_{n_1,n_2-\frac{1}{2}}  + \sum_{n_2\neq n_2^{\prime\prime}} I^z_{n_1+\frac{1}{2},n_2-\frac{1}{2}}\sigmahat^z_{n_1+\frac{1}{2},n_2}\right) \nonumber \\
& + \sum_{n_1\neq n_1^\prime} I^x_{n_1,n_2^\prime} \sigmahat^x_{n_1-\frac{1}{2},n_2^{\prime\prime}}  + \sum_{n_1\neq n_1^{\prime\prime}} I^z_{n_1+\frac{1}{2},n_2^{\prime\prime}-\frac{1}{2}} \sigmahat^z_{n_1,n_2^\prime-\frac{1}{2}} \nonumber \\
& + I^x_{n_1^\prime,n_2^\prime} \left(\prod_{n_1\neq n_1^\prime} \sigmahat^x_{n_1-\frac{1}{2},n_2^{\prime\prime}}\right)\left( \prod_{n_1=1}^{N_1}\prod_{n_2\neq n_2^\prime} \sigmahat^x_{n_1,n_2-\frac{1}{2}}\right)  \nonumber \\
& + I^z_{n_1^{\prime\prime}+\frac{1}{2},n_2^{\prime\prime}-\frac{1}{2}} \left(\prod_{n_1\neq n_1^{\prime\prime}} \sigmahat^z_{n_1, n_2^\prime-\frac{1}{2}}\right)\left(\prod_{n_1=1}^{N_1}\prod_{n_2\neq n_2^{\prime\prime}} \sigmahat^z_{n_1+\frac{1}{2},n_2}\right). \label{eq:H4}
\end{align}

Note that under $\Uhat_4$, every $\Xhat$ and $\Zhat$ term in the toric code model transforms into an individual emergent qubit, except the last two `accumulation' terms in Eq.~\eqref{eq:H4}. Here $\Uhat_{4_1}$ and $\Uhat_{4_2}$ do what $\Uhat_{3_1}$ and $\Uhat_{3_2}$ did on cylinder or sheet, but due to periodic boundary condition on torus, their action results in the accumulation of qubit operators along $n_2^\prime$ and $n_2^{\prime\prime}$ lines. These accumulations are cleared by $\Uhat_{4_3}$ and $\Uhat_{4_4}$, but their remnant inevitably survives as the last two terms in Eq.~\ref{eq:H4} representing the constraints $\prod_{n_1=1}^{N_1}\prod_{n_2=1}^{N_2} \Xhat_{n_1,n_2} =\hat{1} = \prod_{n_1=1}^{N_1}\prod_{n_2=1}^{N_2}\Zhat_{n_1+\frac{1}{2},n_2-\frac{1}{2}}$. Interestingly, if we take $I^x_{n_1^\prime,n_2^\prime}=I^z_{n_1^{\prime\prime}+\frac{1}{2},n_2^{\prime\prime}-\frac{1}{2}}=0$, then for this `punctured' toric code model, we will not get the two accumulation terms! This is very similar to the nearest-neighbour Ising chain under duality transformation; the toric code model on torus is a complex quantum analog of the closed Ising chain, and its punctured version described above is analogous to the open Ising chain.

In $\Uhat^\dag_4\Hhat_4\Uhat_4$, with or without this puncture, we also have two missing qubits at $(n_1^\prime-\frac{1}{2},n_2^{\prime\prime})$ and $(n_1^{\prime\prime},n_2^\prime-\frac{1}{2})$, shown in Fig.~\ref{fig:toric-code}($b$) by a blue and a purple circle. They give rise to the well-known topological degeneracy of four in every conserved sector given by $2(N_1N_2-1)$ emergent qubit quantum numbers $\{z\}$ and $\{x\}$. 
Below we derive an exact expression for all the eigenstates of the toric code model, $\Hhat_4$ (with or without puncture), in the pure density matrix form. 
\begin{eqnarray}
&& \rhohat^{ }_{\mbox{\tiny toric-code}} = \prod_{(n_1,n_2)\neq (n_1^\prime,n_2^\prime)} \left(\frac{\hat{1}+x_{n_1,n_2}\Xhat_{n_1,n_2}}{2}\right) \times \nonumber \\
&& \prod_{(n_1,n_2)\neq (n_1^{\prime\prime},n_2^{\prime\prime})}\left(\frac{\hat{1}+z_{n_1+\frac{1}{2},n_2-\frac{1}{2}}\Zhat_{n_1+\frac{1}{2},n_2-\frac{1}{2}}}{2}\right) \times \nonumber \\
&&\left[\frac{1+\chi\left(u_z \prod_{n_1=1}^{N_1} \sigmahat^z_{n_1-\frac{1}{2},n^{\prime\prime}_2} + u_x \prod_{n_2=1}^{N_2} \sigmahat^x_{n_1^\prime-\frac{1}{2},n_2}\right)}{2} \right] \times \nonumber \\
&&\left[\frac{1+\zeta\left(v_z \prod_{n_2=1}^{N_2} \sigmahat^z_{n_1^{\prime\prime},n_2-\frac{1}{2}} + v_x \prod_{n_1=1}^{N_1} \sigmahat^x_{n_1,n^{\prime}_2-\frac{1}{2}}\right)}{2}\right] 
\label{eq:rho-toric-code}
\end{eqnarray}
Besides the quantum numbers $\{z\}$ and $\{x\}$, the $\rhohat_{\mbox{\tiny toric-code}}$ has in it two other quantum numbers $\chi=\pm1$ and $\zeta=\pm1$ for the two qubits that go missing in $\Uhat_4^\dag\Hhat_4\Uhat_4$. It also depends on two arbitrary real unit vectors $\mathbf{u}=(u_x,u_y,u_z)$ and $\mathbf{v}=(v_x,v_y,v_z)$, which are the quantization axes of these missing qubits at sites $(n_1^\prime-\frac{1}{2},n_2^{\prime\prime})$ and $(n_1^{\prime\prime},n_2^\prime-\frac{1}{2})$ respectively. In Eq.~\eqref{eq:rho-toric-code}, we have taken $u_y=v_y=0$ for some simplicity; one can take them to be nonzero, if so required. 

Notably, the eigenstate given by Eq.~\eqref{eq:rho-toric-code} is a general linear superposition in the fourfold degenerate eigensubspace for a given $\{z\}$ and $\{x\}$. This linear superposition is effected by the quantization axes $\mathbf{u}$ and $\mathbf{v}$ of the two missing qubits. The energy eigenvalues of the toric code model, $E_{\{z\},\{x\}} = \sum_{(n_1,n_2)\neq(n_1^\prime,n_2^\prime)} I^x_{n_1,n_2} x_{n_1,n_2} + \sum_{(n_1,n_2)\neq(n_1^{\prime\prime},n_2^{\prime\prime})} I^z_{n_1+\frac{1}{2},n_2-\frac{1}{2}} z_{n_1+\frac{1}{2},n_2-\frac{1}{2}} + I^z_{n_1^{\prime\prime}+\frac{1}{2},n_2^{\prime\prime}-\frac{1}{2}}\prod_{(n_1,n_2)\neq (n_1^{\prime\prime},n_2^{\prime\prime})} z_{n_1+\frac{1}{2},n_2-\frac{1}{2}} +  I^x_{n_1^\prime,n_2^\prime}\prod_{(n_1,n_2)\neq (n_1^\prime,n_2^\prime)} x_{n_1,n_2} $, do not depend on $\zeta$ and $\chi$; hence the degeneracy of four. 

\section{Quantum Circuits for the Toric Code and Other Eigenstates\label{sec:circuits}} 
Interestingly, the unitary transformations constructed here also provide an exact basis for their implementation as quantum circuits. We note that an operator of the form $(\sigmahat^z_{1}\sigmahat^z_{3})^{\Qhat^x_{2}}$ reduces exactly into a product of two CNOT quantum gates, $(\sigmahat^x_2)^{\Qhat^z_1} (\sigmahat^x_2)^{\Qhat^z_3}$, with a common target qubit at $2$ and two control qubits at $1$ and $3$. From this it immediately follows that the unitary operators like in Eq.~\eqref{eq:U} for the model on trestle, or in Eq.~\eqref{eq:U4} for the toric code model, can all be implemented as quantum circuits using CNOT gates. Below we present the quantum circuits for the eigenstates constructed in this paper. Besides the two-qubit CNOT gates, these circuits also use single-qubit Hadamard gates, and for the toric code eigenstates on torus, a few rotation gates as well. 

\subsection{Circuit for the eigenstates on trestle}
\begin{figure}[t]
\centering
\includegraphics[width=\columnwidth]{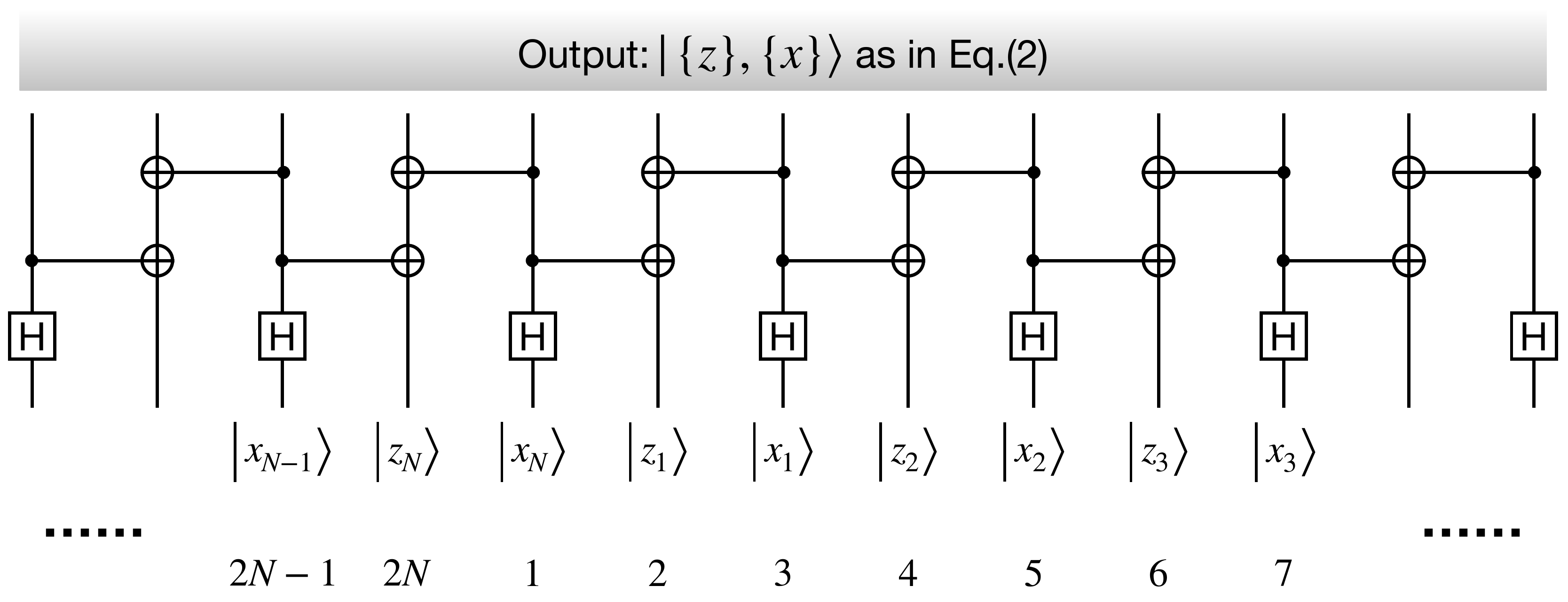}
\caption{Quantum circuit for generating the eigenstate, $\ketzx$, of the model on trestle. Here $\fbox{\sf H}$ stands for Hadamard gate, and a horizontal line connecting a control qubit ($\bullet$) with a target qubit ($\oplus$) denotes a CNOT gate. Input states of the target qubits are given independently by the quantum numbers $\{z\}$ and those of the control qubits by $\{x\}$. It produces the many-qubit state $\ketzx$ of Eq.~\eqref{eq:ketzx}.}
\label{fig:circuit-trestle}
\end{figure}

We can produce the eigenstates of the model on trestle, i.e. $\ketzx$ of Eq.~\eqref{eq:ketzx}, using the quantum circuit shown in Fig.~\ref{fig:circuit-trestle}. It works as follows:
\begin{enumerate}
\item Prepare the input product state, $\prod_{l=1}^{2N}\otimes\ket{\sigma_l}$, with $\sigma_{2n}=z_n$ and $\sigma_{2n+1}=x_n$ for all $n=1$ to $N$. Here (as well as in all the other circuits described later), the input state is prepared in the standard basis, $\ket{+} \equiv \ket{0}$ and $\ket{-} \equiv \ket{1}$, of the $\sigmahat^z$ operators. 
\item Apply Hadamard gates on the odd-numbered qubits.
\item Apply in parallel the CNOT gates first on the pairs of qubits $(1,2)$, $(3,4)$, $\dots (2N-1,2N)$, and then on the pairs $(2,3)$, $(4,5)$, $\dots (2N,1)$. Here the even-numbered qubits are the target qubits and the odd-numbered qubits act as the control qubits.
\end{enumerate}
This circuit returns Eq.~\eqref{eq:ketzx} as output. For a total of $2N$ qubits, it uses $N$ Hadamard gates and $2N$ CNOT gates. The depth of this quantum circuit is 3 for any $N$. A closely related circuit is reported to has been implemented in Ref.~\cite{Bluvstein2022}. 

For the general model $\Hhat_2$ on an arbitrary graph, the $\Uhat_2$ in Eq.~\eqref{eq:U2}, a product of the commuting terms 
$\left[\prod_{n^\prime} \tauhat^z_{(n,n^\prime)}\right]^{\Qhat^x_n} = \prod_{n^\prime} (\sigmahat^x_n)^{\Qhat^z_{(n,n^\prime)}}$, can likewise be implemented 
as a quantum circuit of CNOT gates with qubits on the bonds $\{(n,n^\prime)\}$ acting as controls for the targets at sites $n$. To generate the eigenstate in Eq.~\eqref{eq:ketzx-graph}, 
\begin{itemize}
\item[-] first apply the Hadamard gates on the control (bond) qubits prepared in the input state $\otimes \ket{x_{(n,n^\prime)}}$, and
\item[-] then apply the CNOT gates between these control (bond) qubits and their target (site) qubits prepared in the input state $\otimes \ket{z_n}$.
\end{itemize}
The depth of this circuit scales linearly with the largest coordination number of the target qubits, but not with the total number of qubits. The number of Hadamard gates required is same as the number of bonds, $N_b$, in a graph. The number of CNOT gates required scales linearly with the number of sites, $N_s$, and depends on the average coordination number as well. 

\begin{figure}[b]
\centering
\includegraphics[width=\columnwidth]{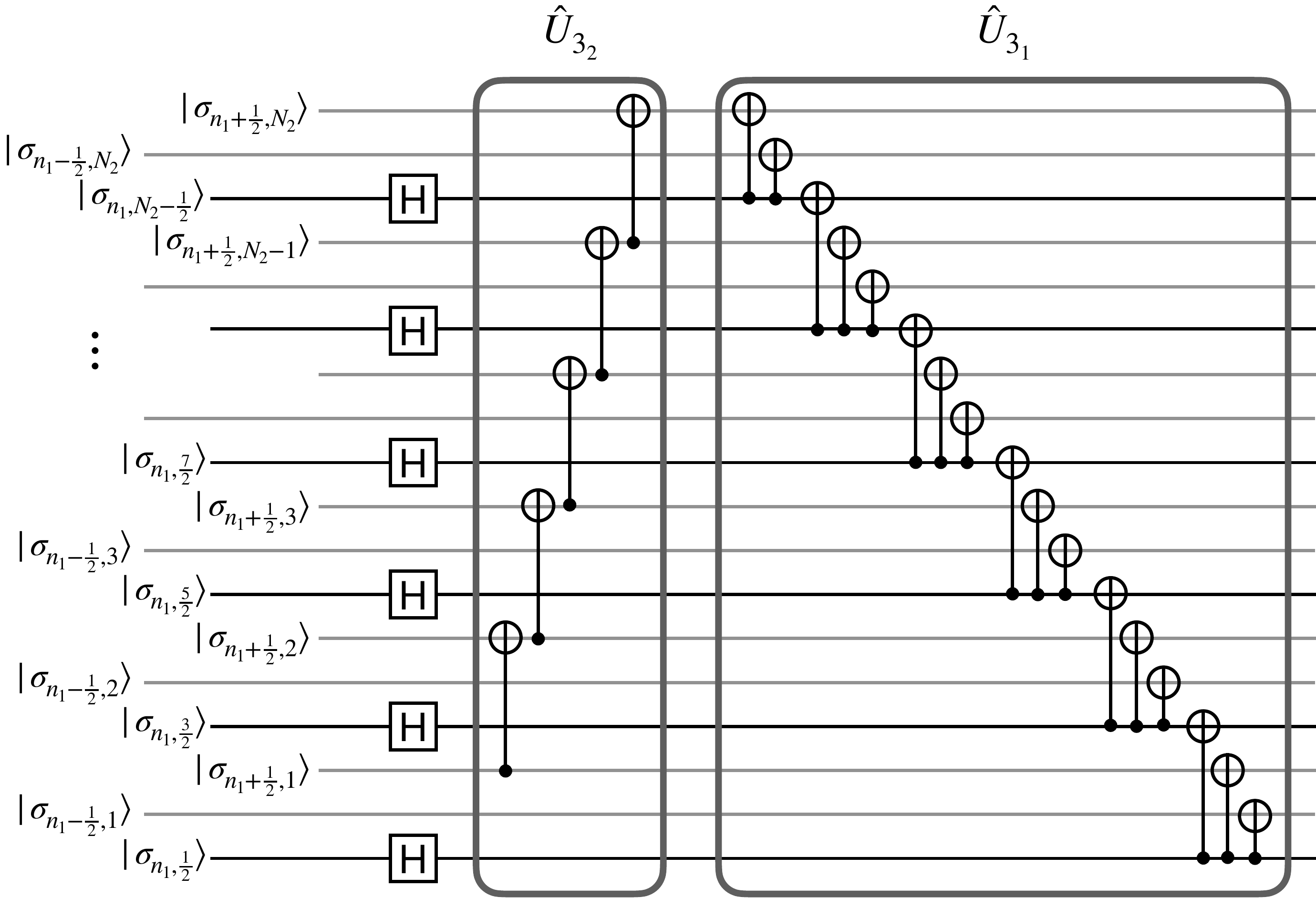}
\caption{Quantum circuit for generating the eigenstates of the toric-code model on cylinder as in Fig.~\ref{fig:cylinder-sheet}$(a)$. Refer to Eqs.~\eqref{eq:H3} and~\eqref{eq:U3} for a clear reading of this diagram. Inside the first box from left, we show the circuit for that part of $\Uhat_{3_2}$ which acts on the qubits (empty circles in Fig.~\ref{fig:cylinder-sheet}) sitting on a fixed $n_1+\frac{1}{2}$ line. The other parts of $\Uhat_{3_2}$ for different values of $n_1$ act in parallel. Inside the second box, we show the circuit for $\Uhat_{3_1}$ for a fixed $n_1$. The $\Uhat_{3_1}$ can be applied first for $n_1=1,3,5\dots$ in parallel and then for $n_1=2,4,6\dots$. In the input state, $\sigma_{n_1+\frac{1}{2},n_2}=z_{n_1+\frac{1}{2},n_2-\frac{1}{2}}$ and $\sigma_{n_1,n_2-\frac{1}{2}}=x_{n_1,n_2}$. A circuit for the planar model in Fig.~\ref{fig:cylinder-sheet}$(b)$ can be constructed similarly. }
\label{fig:circuit-U3}
\end{figure}

\subsection{Circuit for the toric code eigenstates on cylinder}
The quantum circuit for the eigenstates, Eq.~\eqref{eq:ketzx-cylinder}, of the toric-code model on cylinder 
works as follows:
\begin{enumerate}
\item Prepare the input state of all the qubits on constant $n_1$ lines (filled circles) in Fig.~\ref{fig:cylinder-sheet} as a product state $\prod_{n_1=1}^{N_1}\prod_{n_2=1}^{N_2} \otimes \ket{\sigma_{n_1,n_2-\frac{1}{2}}=x_{n_1,n_2}}$. 
\item Prepare the input state of all the qubits on constant $n_2$ lines (empty circles) in Fig.~\ref{fig:cylinder-sheet} as a product state $\prod_{n_1=1}^{N_1}\prod_{n_2=1}^{N_2} \otimes \ket{\sigma_{n_1+\frac{1}{2},n_2}=z_{n_1+\frac{1}{2},n_2-\frac{1}{2}}}$. 
\item Apply Hadamard gates on the qubits prepared in step~1; see Fig.~\ref{fig:circuit-U3}. It requires a total of $N_1N_2$ Hadamard gates. 
\item Apply $\Uhat_{3_2}$ of Eq.~\eqref{eq:U3_2} on the qubits prepared in step~2. The chains of operations for different $n_1$'s act in parallel, but the operations for a given $n_1$ act in series.  The circuit for $\Uhat_{3_2}$ for a fixed $n_1$ is shown in Fig.~\ref{fig:circuit-U3} in the first box. A total of $N_1(N_2-1)$ CNOT gates are required to implement $\Uhat_{3_2}$. 

\item Apply $\Uhat_{3_1}$ as defined in Eq.~\eqref{eq:U3_1}. The strings of operations in $\Uhat_{3_1}$ for different $n_1$'s commute, but successive strings (say, for $n_1$ and $n_1+1$) act commonly on the qubits lying in-between (on $n_1+\frac{1}{2}$ lines). Hence, $\Uhat_{3_1}$ is applied first for $n_1=1,3,5\dots$ in parallel, and then for $n_1=2,4,6\dots$ in parallel. The circuit for a given $n_1$ is shown in Fig.~\ref{fig:circuit-U3}. The total number of CNOT gates required to implement $\Uhat_{3_1}$ is $N_1(3N_2-1)$.  
\end{enumerate}
The total number of CNOT gates required to make this circuit for producing an eigenstate of the model on cylinder is $2N_1(2N_2-1)$. In addition, one requires $N_1N_2$ Hadamard gates. Hence, the net requirement of quantum gates for this circuit scales linearly with the total number of qubits. The depth of this quantum circuit is $4N_2-1$, which is sub-extensive and scales linearly with the number of qubits along the open (i.e. $n_2$) direction of the cylinder. A quantum circuit for producing the eigenstates of the planar code in Fig.~\ref{fig:cylinder-sheet}$(b)$ can be constructed exactly in the same manner. 

\subsection{Circuit for the toric-code eigenstates on `torus'}
 Below we present an exact quantum circuit for the unitary transformation $\Uhat_4$, Eq.~\eqref{eq:U4}, that produces toric code eigenstates, Eq.~\eqref{eq:rho-toric-code}, on a torus by acting on independent qubits. It goes as follows:

\begin{figure}[t]
\centering
\includegraphics[width=\columnwidth]{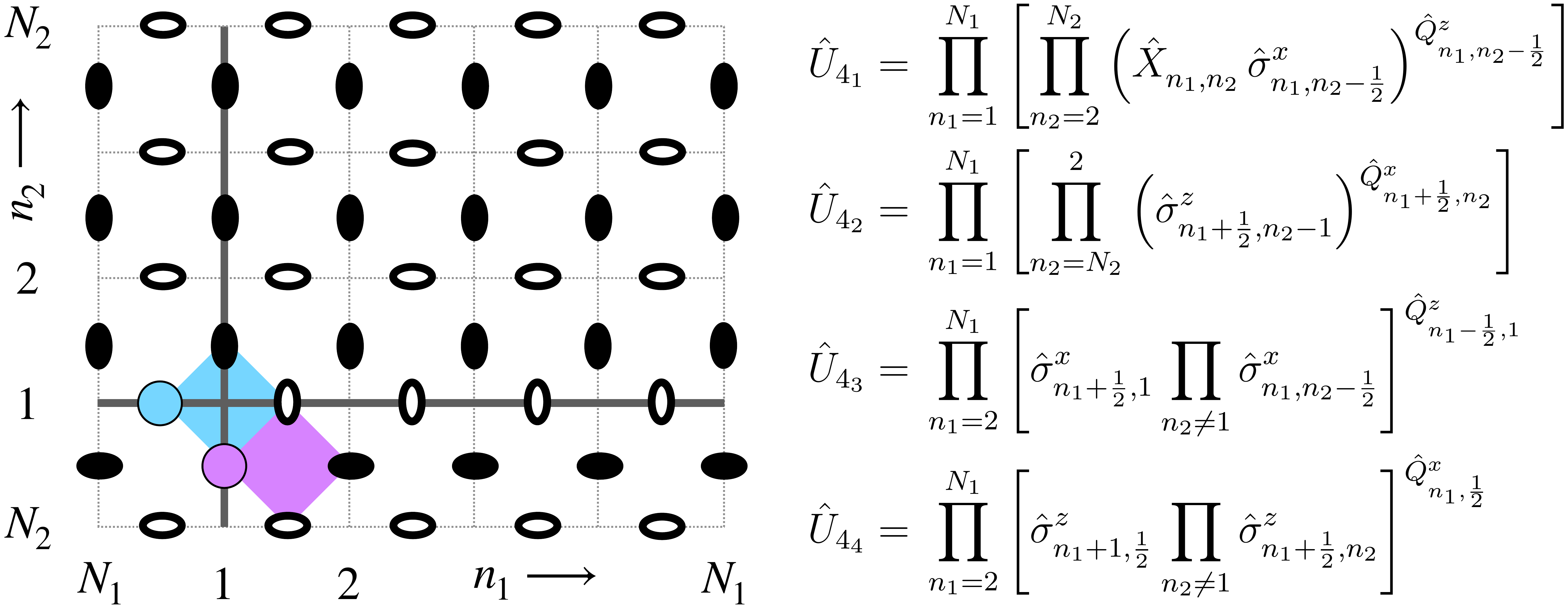} \\ 
\includegraphics[width=\columnwidth]{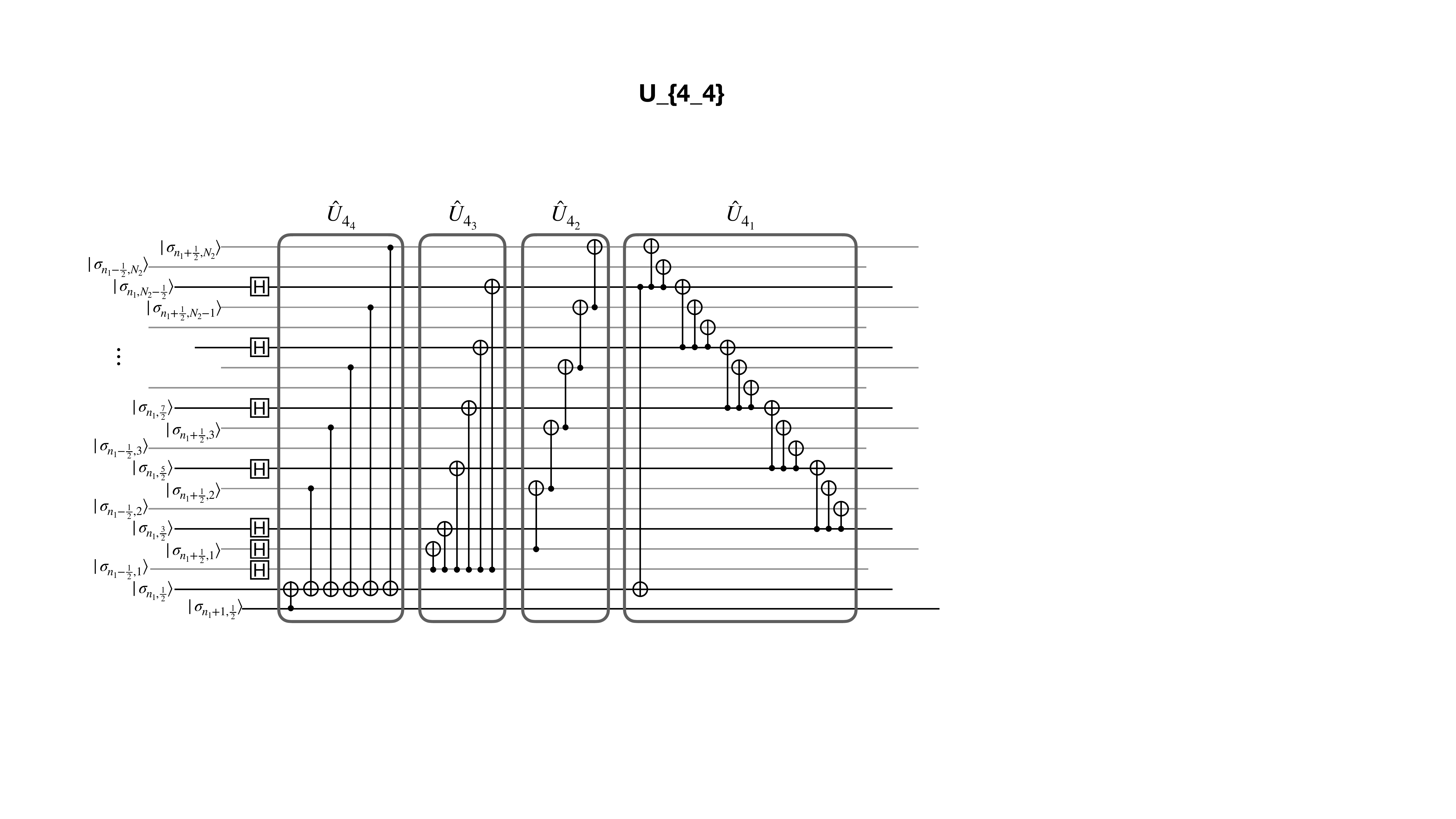} 
\caption{Quantum circuit for generating the toric-code eigenstates on torus. $Top$: A particular version of Fig.~\ref{fig:toric-code}$(b)$ and Eqs.~\eqref{eq:U4} for $n_1^\prime=n_1^{\prime\prime}=1=n_2^\prime=n_2^{\prime\prime}$. Figure on top-left presents independent qubits with their input state for the $\Uhat_4=\Uhat_{4_1}\Uhat_{4_2}\Uhat_{4_3}\Uhat_{4_4}$ constituted by the operators on top-right. $Bottom$: Circuit for $\Uhat_4$. Only a part each of the circuits for $\Uhat_{4_4}$, $\Uhat_{4_3}$, $\Uhat_{4_2}$ and $\Uhat_{4_1}$, for fixed $n_1$, is shown. Different parts of $\Uhat_{4_4}$ for $n_1$ going from $N_1$ to  $2$ act in series; the same is true for $\Uhat_{4_3}$. The $\Uhat_{4_4}$ and $\Uhat_{4_3}$ act in parallel, followed by $\Uhat_{4_2}$ followed by $\Uhat_{4_1}$. Different terms of $\Uhat_{4_2}$ for different $n_1$'s act in parallel. The $\Uhat_{4_1}$ is applied first for $n_2=2,4,6\dots$ in parallel, and then for $n_1=1,3,5\dots$ in parallel.}
\label{fig:circuit-toric-code}
\end{figure}

\begin{enumerate}
\item For simplicity, take $n_1^\prime=n_1^{\prime\prime}=1=n_2^\prime=n_2^{\prime\prime}$; refer to Fig.~\ref{fig:toric-code}$(b)$ for the notation, and Fig.~\ref{fig:circuit-toric-code} for this particular design. In this case, the two qubits, that go missing in the toric code model transformed under $\Uhat_4$, are located at $(\frac{1}{2},1)$ and $(1,\frac{1}{2})$. 

\item  Prepare the qubits on the $n_2=1$ horizontal line (empty vertical ovals in Fig.~\ref{fig:circuit-toric-code}) in the product state $\prod_{n_1=1}^{N_1}\otimes \ket{\sigma_{n_1-\frac{1}{2},1}=x_{n_1,1}}$. Here $\sigma_{\frac{1}{2},1}=x_{1,1}\equiv \chi$ is the quantum number of the `missing' qubit at $(\frac{1}{2},1)$; refer to Eq.~\eqref{eq:rho-toric-code}.  

\item Prepare the qubits (empty horizontal ovals) on all the other horizontal lines for $n_2=2, 3, \dots, N_2$ in the product state $\prod_{n_2=2}^{N_2}\prod_{n_1=1}^{N_1}\otimes \ket{\sigma_{n_1+\frac{1}{2},n_2}=z_{n_1+\frac{1}{2},n_2-\frac{1}{2}}}$.  

\item  Prepare the qubits, filled horizontal ovals in Fig.~\ref{fig:circuit-toric-code}, sitting on the horizontal line between $n_2=N_2$ and 1 in the product state $\prod_{n_1=1}^{N_1}\otimes \ket{\sigma_{n_1,\frac{1}{2}}= z_{n_1+\frac{1}{2},\frac{1}{2}}}$. Here $\sigma_{1,\frac{1}{2}} = z_{\frac{3}{2},\frac{1}{2}}\equiv\zeta$ is the quantum number of the other `missing' qubit at $(1,\frac{1}{2})$.  

\item Prepare the remaining qubits, filled vertical ovals, in the product state $\prod_{n_2=2}^{N_2}\prod_{n_1=1}^{N_1}\otimes \ket{\sigma_{n_1,n_2-\frac{1}{2}}= x_{n_1,n_2}}$. 

\item Apply Hadamard gates to the qubits prepared in steps 2 and 5, except the `missing' qubit at $(\frac{1}{2},1)$.

\item Apply arbitrary rotations to the two `missing' qubits at $(1,\frac{1}{2})$ and $(\frac{1}{2},1)$. Applying rotation gate ${\sf R}_y(\theta_u)$ on the qubit at $(\frac{1}{2},1)$ and ${\sf R}_y(\theta_v)$ on the qubit at $(1,\frac{1}{2})$ produces the toric-code eigenstate with $\mathbf{u}=(\sin{\theta_u},0,\cos{\theta_u})$ and $\mathbf{v}=(\sin{\theta_v},0,\cos{\theta_v})$~\footnote{By not applying any rotation to the two missing qubits, the toric code eigenstate produced would have $\mathbf{u}=\mathbf{v}=(0,0,1)$. Applying Hadamard gate (without these rotations) to the missing qubit at $(\frac{1}{2},1)$, like the other qubits prepared in step~2, would give the toric code eigenstate with $\mathbf{u}=(1,0,0)$ and $\mathbf{v}=(0,0,1)$. By playing with the input states of the two missing qubits, one can generate different states within a conserved eigensubspace.}.

\item Apply $\Uhat_{4_4}$. For each $n_1$ going sequentially from 2 to $N_1$, it employs $N_2$ CNOT gates with a common target at $(n_1,\frac{1}{2})$, a control qubit at $(n_1+1,\frac{1}{2})$ and $N_2-1$ control qubits at $(n_1+\frac{1}{2},n_2)$ for all $n_2\neq 1$. Refer to Fig.~\ref{fig:circuit-toric-code} for the $\Uhat_{4_4}$ and its circuit. It requires a total of $(N_1-1)N_2$ CNOT gates. 

\item Apply $\Uhat_{4_3}$. For each $n_1$ going sequentially from 2 to $N_1$, it employs $N_2$ CNOT gates with a common control at $(n_1-\frac{1}{2},1)$, a target at $(n_1+\frac{1}{2},1)$ and $N_2-1$ target qubits at $(n_1,n_2-\frac{1}{2})$ for all $n_2\neq 1$. See the corresponding circuit in Fig.~\ref{fig:circuit-toric-code}. It too requires $(N_1-1)N_2$ CNOT gates. Since $\Uhat_{4_4}$ and $\Uhat_{4_3}$ act on mutually exclusive subsets of qubits, the two circuits act in parallel. 

\item Apply $\Uhat_{4_2}$. It is a product of strings that act in parallel for different values of $n_1$, but the operations for a given $n_1$ acts in series along $n_2$. See Fig.~\ref{fig:circuit-toric-code}. It requires a total of $N_1(N_2-1)$ CNOT gates. 

\item Apply $\Uhat_{4_1}$. It is a product of strings of operations, which for different values of $n_1$ commute, but the operations within a string for a fixed $n_1$ act in series along $n_2$ as shown in Fig.~\ref{fig:circuit-toric-code}. Since the neighbouring strings (for $n_1$ and $n_1\pm1$) act on common qubits (along $n_1\pm\frac{1}{2}$ lines), the strings for $n_1= 2, 4, 6 \dots$ having no common qubits can be implemented in parallel,  and then the same is done for $n_1= 1, 3, 5 \dots$.  The total number of CNOT gates required to implement $\Uhat_{4_1}$ is $3N_1(N_2-1)$. 
\end{enumerate}
The depth of this circuit  is dictated by the depth of the circuits for $\Uhat_{4_4}$ and $\Uhat_{4_3}$, which is $(N_1-1)N_2$; it scales linearly with the total number of qubits; $\Uhat_{4_2}$ and $\Uhat_{4_1}$ further add to it a sublinear depth of about $7(N_2-1)$. Hence, the depth of this quantum circuit for the toric-code eigenstates on torus is an extensive number ($\sim N_1N_2$), in notable contrast with the sub-extensive ($\sim N_2$) depth of the circuit on cylinder (or sheet).

\section{Summary\label{sec:sum}}

Inspired by the toric code model, we in this paper have devised unitary transformations which exactly reduce a class of models, including the toric code model, into independent emergent qubits. We demonstrated the basic idea through a one-dimensional construction on a trestle and its generalization on arbitrary graphs, realizing exact quantum paramagnetic eigenstates with free multipolar moments. We rigorously transformed the toric code model on a torus, cylinder and sheet into independent emergent qubits, and derived their eigenstates exactly for arbitrary interaction strengths. 
We further turned these unitary transformations exactly into quantum circuits that can produce the toric code and other eigenstates on quantum processors. The depth of the circuit for toric code eigenstates on torus grows linearly with the total number of qubits, as compared to the sublinear growth on cylinder or sheet. Current experimental realizations of the toric code are reported to be on planar geometries~\cite{Roushan2021,Liu2019,Song2018}, except a recent one on torus~\cite{Bluvstein2022}, and they concern the ground state in a particular form. The quantum circuit we have devised here can generate any toric code eigenstate with a freedom of superposition offered by the two missing qubits on torus.

\begin{acknowledgments}
The author thanks Matthew Fisher for a valuable discussion during a recent visit to KITP Santa Barbara which was supported in part by the National Science Foundation under Grant No. NSF PHY-1748958. 
\end{acknowledgments}

\bibliography{references}

\end{document}